\documentclass[10pt,showpacs,aps,prl,twocolumn,preprint,superscriptaddress]{revtex4}
\usepackage[pdftex]{graphicx}
\usepackage{mathrsfs}
\usepackage{amsmath}
\usepackage{amssymb}
\usepackage{amsfonts}
\usepackage{ae}
\usepackage{units}
\usepackage{appendix}

\begin{document}
%
\title{Parametric excitation of a dc shuttle current via spontaneous symmetry breaking}

\author{Milton E. Pe\~na-Aza}
\affiliation{\small Department of Physics, University of Gothenburg, SE-412 96 G\"oteborg, Sweden}
\author{Alessandro Scorrano}
\affiliation{\small Department of Mechanical and Aerospace Engineering, Sapienza University of Rome, Via Eudossiana 18, 00184 Rome, Italy}
\author{Leonid Y. Gorelik}
\affiliation{\small Department of Applied Physics, Chalmers University of Technology, SE-412 96 G\"oteborg, Sweden}
\date{\today}
%
\begin{abstract}

We investigate theoretically the dynamics of a spatially symmetric shuttle-system subjected to an ac gate voltage. We demonstrate that in such a system parametric excitation gives rise to mechanical vibrations when the frequency of the ac signal is close to the eigenfrequency of the mechanical subsystem. These mechanical oscillations result in a dc shuttle current in a certain direction due to spontaneous symmetry breaking. The direction of the current is defined by the phase shift between the ac gate voltage and the parametrically excited mechanical oscillations. Dependance of the  shuttle current on the dc gate voltage is also analyzed.

\end{abstract}

\maketitle

Fifteen years ago, a novel form of electron transport based on the mechanical vibrations of a metallic nanoparticle coupled to two electrodes via elastic molecular links was proposed by Gorelik \textit{et al.} \cite{Gorelik}. Since then, the shuttle mechanism of transport has been a subject of intensive experimental and theoretical research \cite{gorelik2,Shekhter,armour,azuma,koenig,mosalenko,kim,azuman}. Different architectures ranging from the original double-junction system to transistor-like configurations comprising a gate electrode have been studied \cite{isacsson}.  
The main feature of the orthodox shuttle structures is that a constant potential difference, applied between two fixed electrodes, leads to a dynamical instability that causes the metal nanoparticle oscillate. In this regime, the resonator attains a stable orbit and a sustained current induced by the voltage drop between the electrodes. This current is proportional to the frequency oscillation of the nano-oscillator \cite{Gorelik}.

In the present work, we investigate the possibility to generate shuttle transport between two electrodes at the same electro-chemical potential. In this scheme, we demonstrate that, despite the lack of a bias voltage, a shuttle dc current can still be detected. This charge transport is achieved by applying an ac voltage to a gate electrode which controls the electronic population of a metallic island and, in this form, also the stiffness of the resonator resulting in a parametric mechanical instability at the resonant frequency. This constitutes a new archetype of electron shuttle in which the symmetry breaking effect (direction of the shuttle transportation) does not rely on the presence of any bias voltage. In the phenomena under consideration, the shuttle current is controlled by the phase shift between the mechanical vibrations and gate voltage oscillations. We will show that in this scenario, two different values of the phase shift, which differ from each other by $\pi$, can correspond to a regime of sustained oscillations. The occurrence of these values for the phase shift depends, in particular, on the initial conditions and, as a result, spontaneous symmetry breaking takes place.

In the context of our work, it is worth mentioning that parametric excitation in nanoelectromechanical systems (NEMS) has been also considered in Refs.~\cite{karabalin,kenig,kenig1,midtvedt}

\begin{figure}
\includegraphics[width=8.5cm]{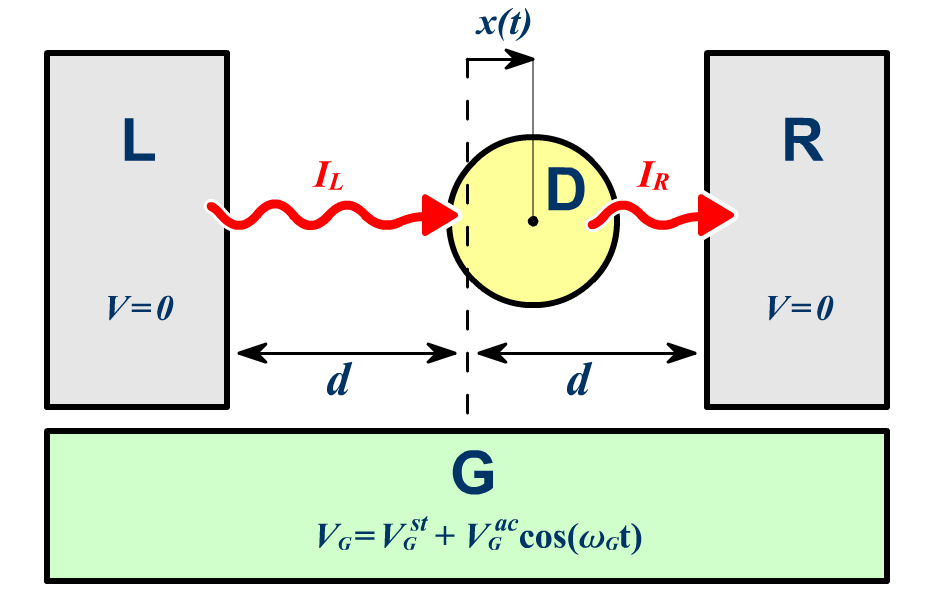}
\caption{(Color online) Schematic diagram of the three-terminals shuttle device investigated in this work. A quantum dot $D$ can oscillate between two metallic leads $L,R$ fixed at zero voltage. The dot is capacitevely coupled to a gate electrode $G$, to which a voltage $V_{g}$ is applied. Electron tunneling takes place between the dot and the leads. $I_{L}$ and $I_{R}$ are the currents between the left and right leads $L,R$ and the dot $D$.}
\label{shuttle}
\end{figure}

To describe the new shuttling mechanism, we consider a system schematically depicted in Fig.~\ref{shuttle} 
where a quantum dot $(D)$ is connected via elastic links to the left $(L)$ and right $(R)$ electrodes. The characteristic dimension of the system is $d$ and the mechanical degree of freedom of the dot is $x(t)$. The voltage over the left and right electrodes is kept null, \textit{i.e.}, $V_{L}=V_{R}=0$, while a signal $V_{G} = V_{G}^{st} + V_{G}^{ac}\cos(\omega_{G}t)$ is applied to the gate (G).

In our description, we focus on the single-electron regime, where the maximum occupancy of the dot is one electron due to Coulomb blockade of tunneling. In addition, we also consider the case in which the mechanical vibration frequency of the dot is very low in comparison to the tunneling rates between the moving quantum dot and electrodes. In this framework, the electronic state of the dot can be described by the average occupation of the dot $n(t)$ and the time evolution of this quantity can be written as
\begin{equation}
\label{masterequation}
\dot{n}=[\Gamma_{L}(x)+\Gamma_{R}(x)](f-n)\,.
\end{equation}
Here, the position-dependent tunneling rates between the left/right electrodes and the dot are  $\Gamma_{L,R}(x)=\Gamma_{0}e^{\mp x/\lambda}$ with $\lambda$ the tunneling length. In addition,  $f=\bigl[1+e^{E_{D}/k_{B}T}\bigr]^{-1}$ is the Fermi-Dirac thermal distribution, where $E_{D}=e^{2}/2C_{T}(x)-e\beta V_{G}$ is the electrostatic energy of the charged dot, $C_{T}(x)$ is the total capacitance of the system, and $\beta \approx 1$ is the transmission coefficient \cite{prl_comment}. Considering the symmetric case $C_{T}(x)=C_{T}(-x)$ and small displacements of the dot, $x \ll d$, we can rewrite $E_{D}=E_{0}-\epsilon e^{2}x^{2} - eV_{G}$, with $E_{0}=e^{2}/2C_{T}(0)$ and $\epsilon \approx d^{-3}$ is a positive parameter.

To consider the mechanical part of the system, we model the quantum dot as a single degree-of-freedom oscillator with eigenfrequency $\omega_{0}$ subjected to an electrostatic force $F_{D}$ induced by the image charges on the leads. Having in mind the quasi-adiabatic condition introduced before, $\omega_{0} \ll \Gamma_{L,R}$, we can make use of the mean field approximation $F_{D}=\langle F_{D}\rangle=-\langle \partial E_{D}/\partial x \rangle = 2\epsilon e^{2}n(t)x$. Therefore, the dynamics of the system is described by the following dimensionless and coupled ordinary differential equations (ODEs),
\begin{subequations}
\label{eq1}
\begin{equation}
\frac{d^2\xi}{d\tau^2}+Q^{-1}\frac{d\xi}{d\tau}+\xi=\tilde{\epsilon}n(\tau)\xi\,\label{eq1a},
\end{equation}
\begin{equation}
\frac{d n}{d\tau}=2\frac{\Gamma_{0}}{\omega_{0}}\cosh(\frac{d}{\lambda}\xi)[f(\xi,\tau)-n(\tau)]\,\label{eq1b},
\end{equation}
\text{where}
\begin{equation}
f(\xi,\tau)=\frac{1}{1+e^{-[\alpha \xi^{2}+\nu_{st}+\nu_{ac}\cos(\tilde{\omega}\tau)]}}.\,\label{eq1c}
\end{equation}
\end{subequations}
Here, $\tau=\omega_{0} t$, $\xi(t)=x(t)/d$, $\tilde{\epsilon}=2e^{2}\epsilon/m\omega_{0}^{2}$, $\alpha=\epsilon e^{2}d^{2}/k_{B}T$, $\nu_{st}=(-E_{0}+eV_{G}^{st})/k_{B}T$, $\nu_{ac}=eV_{G}^{ac}/k_{B}T$,  $\tilde{\omega}=\omega_{G}/\omega_{0}$, and $Q$ is the oscillator quality factor.

The average dimensionless current through the system can be calculated as:

\begin{align}
 \tilde{I} &=-\lim_{\tilde{T}\to\infty}\frac{\Gamma_{0}}{\tilde{T}\omega_{0}}\int^{\tilde{T}}_{0}d\tau\sinh(\frac{d}{\lambda}\xi)[f(\xi,\tau)-n(\tau)]\,, \notag \\ 
&=\lim_{\tilde{T}\to\infty}\frac{d}{2\tilde{T}\lambda}\int^{\tilde{T}}_{0}d\tau\frac{\dot{\xi}(\tau)n(\tau)}{\cosh^{2}(\tau)}\,. \label{current2}
\end{align}
Here, in writing Eq.~\eqref{current2} we use Eq.~\eqref{eq1b} and $\tilde{T}=2\pi/\tilde{\omega}$. From this expression one can see that a dc current between the leads is defined by the correlations between the velocity and population of the dot. To find these correlations one should analyze the dynamical system described by Eqs.~\eqref{eq1}. 

A formal solution of Eq.~\eqref{eq1b} is given by,
\begin{equation}
\label{eq2}
n(\tau)=\frac{2\Gamma_{0}}{\omega_{0}}\int_{-\infty}^{\tau}d\tau'e^{-\frac{2\Gamma_{0}}{\omega_{0}}\smallint^{\tau}_{\tau'}dt''\cosh(\tilde{d}\xi)}\cosh(\tilde{d}\xi)f(\xi,\tau').
\end{equation}
Then, by substituting Eq.~\eqref{eq2} into Eq.~\eqref{eq1a} one can, in principle, obtain the exact dynamics of the system by solving the resulting integro-differential equation. However, exploiting the smallness of the parameters $\omega_{0}/\Gamma_{0}, \tilde{\epsilon},Q^{-1}\ll 1$, a perturbative analysis can be employed. In doing so, in the leading order of the parameter $\omega_{0}/\Gamma_{0}$, Eq.~\eqref{eq1b} reduces to $n(\tau)=f(\xi,\tau)$ and the oscillator equation of motion, Eq.~\eqref{eq1a}, takes the form:
\begin{equation}
\label{simple}
\frac{d^2\xi}{d\tau^2}+Q^{-1}\frac{d\xi}{d\tau}+\xi=\tilde{\epsilon}f(\xi,\tau)\xi\,.
\end{equation}
Considering the second harmonic in the Fourier expansion of $f(\xi,\tau)$ in Eq.~\eqref{simple}, one can find that the system experiences a parametric mechanical instability at the driving frequency $\tilde{\omega}\approxeq 1$. As this takes place, in the stationary regime, the time evolution of the dot position $\xi(t)$, in the leading order of small parameters $\tilde{\epsilon},Q^{-1}\ll 1$, is given by the expression: 

\begin{equation}
\label{eq3}
\xi(\tau)=A\cos(\tilde{\omega}\tau + \chi)\,.
\end{equation}
Here, the quantities $A \equiv \sqrt{2E}$ and $\chi$ are the stationary dimensionless amplitude of the dot and the phase shift between the mechanical and gate voltage oscillations, respectively. The oscillation amplitude, generated by the parametric instability,  is bounded by nonlinearities and it is controlled by the parameter $\alpha$.

Turning back to the charge transport, on the accuracy of $\omega_{0}/\Gamma_{0}$, the average population of the dot becomes $n(\tau)=f(A\cos(\tilde{\omega}\tau + \chi),\tau)$ using the ansatz in Eq.~\eqref{eq3}. Thus, the expression for the average current of the system in Eq.~\eqref{current2} can be recasted as:

\begin{subequations}
\label{current3}
\begin{equation}
\tilde{I}=\frac{d}{2\lambda}\int^{A}_{-A}\frac{d\xi}{\cosh^{2}(\xi)}\frac{\sinh(b\vert \dot{\xi}(\xi) \vert\sin(\chi))}{[\cosh(a(\xi))+\cosh(b\vert \dot{\xi}(\xi) \vert\sin(\chi))]},
\end{equation}
\text{with}
\begin{equation}
a(\xi)=\alpha \xi^{2}+\nu_{st}+b\tilde{\omega}\xi\cos(\chi)\,,\quad b=\nu_{ac}/\tilde{\omega}A.
\end{equation}
\end{subequations}
Here, $\vert \dot{\xi}(\xi) \vert$ is the modulus of the dot velocity as a function of its position given by the relation $\tilde{\omega}^2\vert \dot{\xi}(\tau)\vert^{2}+\vert\xi(\tau)\vert^2=2E$. From Eq.~\eqref{current3}, for small amplitudes and $\tilde{\omega}\backsim 1$, the current of the system is given by
\begin{subequations}
\label{current4}
\begin{equation}
 I=\omega_{0}e\varkappa(d/\lambda)A
\end{equation}
\text{with}
\begin{equation}
 \varkappa=\frac{\sinh(\nu_{ac}\sin(\chi))}{\cosh(\nu_{st}+\nu_{ac}\cos(\chi))+\cosh(\nu_{ac}\sinh(\chi))}
\end{equation}
\end{subequations}

\begin{figure}
\includegraphics[width=8.5cm]{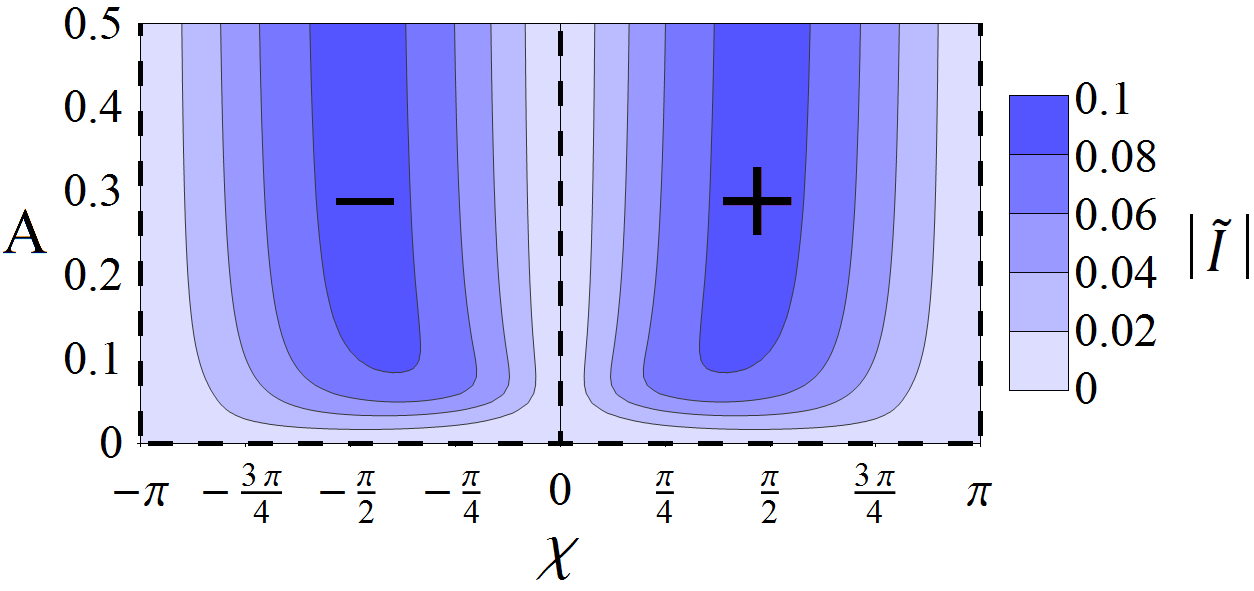}
\caption{(Color online) Contour plot for the dimensionless current $\vert\tilde{I}\vert$ as a function of the amplitude $A$ and phase $\chi$. The plot is symmetric with respect to the phase and the sign of the current depends on the sign of the phase. The current vanishes at points (black dash lines) $A=0$ and/or $\chi=\pm n\pi, n=0,1,2,\ldots$, current is maximum around $\chi=\pm \pi/2$.} 
\label{corriente}
\end{figure}

From Eqs.~\eqref{current3} and \eqref{current4}, we can conclude that current is only attained at a non-zero dot oscillation amplitude $A$ while its direction (symmetry breaking) is controlled by the phase difference $\chi$. This behavior is visulized in Fig.~\ref{corriente}, which is a contour plot of the dimensionless average current as a function of the amplitude and phase. From the plot it becomes clear that the sign of the current follows the sign of the phase. We also notice that the electronic transport ceases when the amplitude is exactly zero $A\equiv 0$ and/or the phase $\chi=\pm n\pi$, with $n=0,1,2\ldots$, points highlighted in the phase space in dashed black lines. The plot also indicates that the current is maximum around $\chi=\pm\pi/2$.

Once we have studied the shuttle transport current of the system and investigated how it arises from a parametric excitation. We proceed by analyzing the stationary amplitude-phase characteristics of the dot as a function of the parameters $\nu_{dc}$ and $\nu_{ac}$, the dimensionless applied voltages. To proceed in this direction, we will assume that $A$ and $\chi$ vary slowly in time. Then, substituting Eq.~\eqref{eq3} into Eq.~\eqref{eq1c} and the resulting expression in Eq.~\eqref{eq1a}, after averaging over the fast oscillations one obtains the following coupled differential equations for $E(\tau)$ and $\chi(\tau)$,
\begin{subequations}
 \label{eq4}
\begin{align}
\dot{E}&=\frac{\partial \mathscr{H}}{\partial \chi} - Q ^{-1}E,\label{eq4a}\\
\dot{\chi}&=-\frac{\partial \mathscr{H}}{\partial E} .\label{eq4b}
\end{align}
\text{Here, $\mathscr{H}$ is the generating Hamiltonian function,}
\begin{multline}
\label{hamiltonian}
\mathscr{H}(E,\chi)=-(1-\tilde{\omega})E +\\
+\frac{\tilde{\epsilon}}{2\alpha \pi}\int^{\pi}_{-\pi}d\theta\ln\{1+e^{[\alpha E\cos^{2}(\theta)+\nu_ {st}+\nu_{ac}\cos(\theta-\chi)]}\}.
\end{multline}
\end{subequations}
\begin{figure}
\includegraphics[width=7cm]{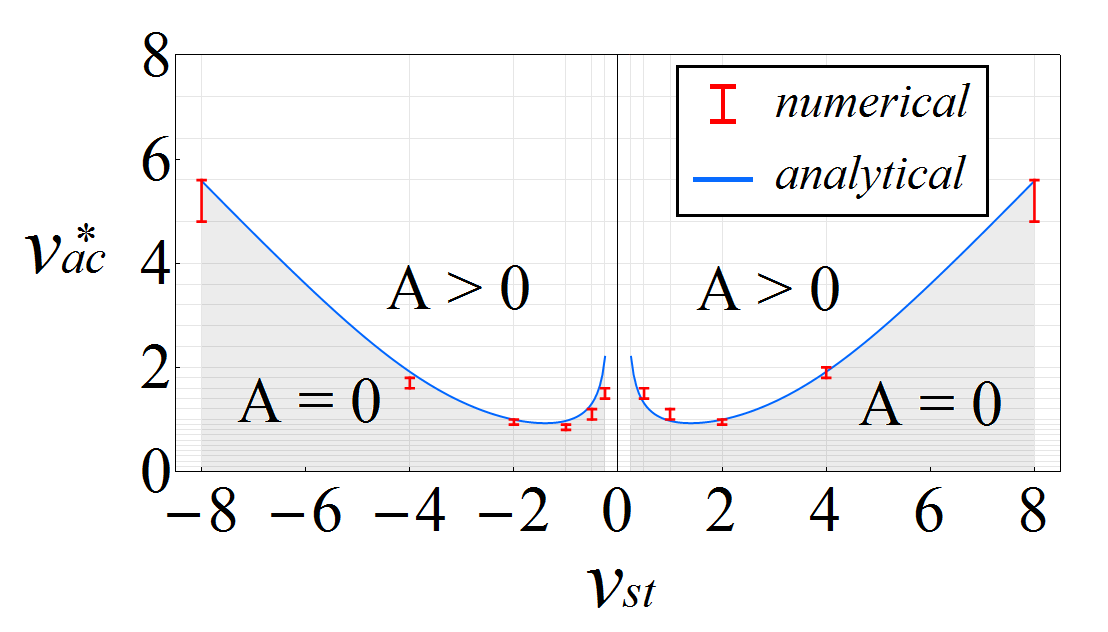}
\caption{(Color online) Threshold ac voltage $\nu_{ac}^{*}$ as a function of the stationary dc voltage $\nu_{st}$. Red bars correspond to results form numerical integration of Eqs.~\eqref{eq1} while blue lines refer to the analytical perturbative scheme carried out in Eq.~\eqref{instability}. The plot is calculated for a gold quantum dot of radius $r=$~\unit[4]{nm} and mass $m=$~\unit[5$\times 10^{-21}$]{kg} with $\omega_{0}=$~\unit[10]{GHz}, $Q=$~\unit[1000], $d\sim$~\unit[2]{nm}, $\lambda=$~\unit[0.1]{nm}, $\Gamma_{0}=$~\unit[100]{GHz}, $\omega_{G}=$~\unit[10]{GHz} and $T=$~\unit[10]{K}. Consequently, $\tilde{\epsilon}=0.1$, $\alpha= 837$, $\tilde{\omega} = 1$. }
\label{threshold}
\end{figure}

Further analyses from Eq.~\eqref{eq4} indicate that given a certain stationary dc voltage $\nu_{st}$ there exists a critical amplitude of the ac voltage $\nu_{ac}^{*}$ above which the parametric excitation of the mechanical vibration gives rise, see Fig.~\ref{threshold}. This fact can be understood in the frame of a perturbative analysis for small $E$. In this regime, we find that the instability criterion due to the parametric resonance is given by

\begin{equation}
\frac{1}{{Q\tilde{\epsilon}}}\leq \frac{\sin(2\chi_{st})}{4\pi}\int^{\pi}_{-\pi}d\theta\cos(2\theta)\tanh(\frac{\nu_{st}+\nu_{ac}\cos(\theta)}{2}).
\label{instability}
\end{equation}
From Eq.~\eqref{instability} one can also notice that the right hand side of the inequality vanishes in the limit $\nu_{st}\to 0$ and the system can not be parametrically excited. In the instability criterion $\chi_{st}$ is defined through the equation,

\begin{equation}
\label{statico}
\left. \frac{\partial{\mathscr{H}(E,\chi_{st})}}{\partial{E}} \right|_{E=0}=0\,.
\end{equation}

By substituting $\chi_{st}$ into Eq.~\eqref{instability} it is possible obtain a relation $\nu_{ac}^{*}=\nu_{ac}^{*}(\nu_{st})$ for the critical ac voltage as a function of the stationary dc voltage. This threshold ac voltage is shown in Fig.~\ref{threshold}, where red bars correspond to results coming from the numerical integration of Eqs.~\eqref{eq1} while blue lines refer to the analytical perturbative scheme carried out in Eq.~\eqref{instability}. The agreement between the two approaches is noticeable.

In the remainder of this work, we will be mainly focusing on the case of exact resonance, $\tilde{\omega}=1+(\tilde{\epsilon}/4\pi)\int_{-\pi}^{\pi}d\theta[e^{-\nu_{st}}+e^{\nu_{ac}\cos(\theta-\chi)}]^{-1}$, and would like to discuss in detail the outcomes from the stability analysis. Due to the periodicity of the generating Hamiltonian function, $\mathscr{H}(E,\chi)= \mathscr{H}(E,\chi+\pi)$, the stationary solutions of Eq.~\eqref{eq4} come in pairs: to any solution $P_{i}=\{\chi_{i},A_{i}\}$ corresponds a conjugated solution $\bar{P}_{i}=\{\chi_{i}+\pi,A_{i}\}$. Computational analyses have shown that for $\nu_{ac}(\nu_{st})<\nu_{ac}^{*}(\nu_{st})$ the system in Eq.~\eqref{eq4} possesses four formal stationary points $P_{1}=\{\pi/4,0\}$, $P_{2}=\{3\pi/4,0\}$ and their conjugates (for $\nu_{st}>0$, $P_{1}$($\bar{P}_{1}$) and $P_{2}$($\bar{P}_{2}$) are stable and unstable, respectively. For $\nu_{st} < 0$, viceverse. If $\nu_{ac}(\nu_{st})>\nu_{ac}^{*}(\nu_{st})$, two stationary stable points $P_{3}=\{\chi_{3}, A_{3}\}$ and  $\bar{P}_{3}$ appear on the phase diagram while the points $P_{1}$ and $P_{2}$ (and their conjugates) become unstable.

\begin{figure}
\includegraphics[width=7.5cm]{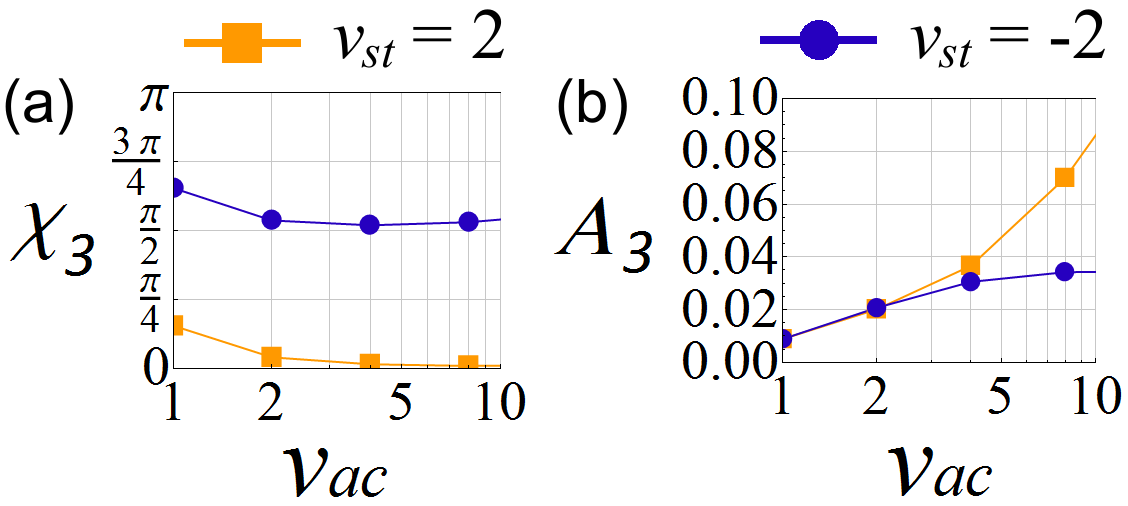}
\caption{(Color online) Stationary phase (a) and amplitude (b) of the system as a function of the applied dc voltage $\nu_{st}=2$ (orange square) and $\nu_{st}=-2$ (blue circles) for different dc voltages. The phase is almost zero for the positive dc voltage while it is non-vanishing for the negative dc voltage. Current transport is more feasible at negative dc voltages. The plot is calculated for $\tilde{\epsilon}=0.1$, $\alpha= 837$, $\tilde{\omega} = 1$. }
\label{stability}
\end{figure}

In Fig.~\ref{stability}, the phase shift and amplitude of the stable solution $P_{3}=\{\chi_{3},A_{3}\}$ are shown as a function of the applied ac voltage for two different dc voltages: $\nu_{st}=2$ (blue circles) and $\nu_{st}=-2$ (orange squares). From this graph one can notice that for the positive dc voltage, the phase shift and, as a consequence, the average current (see Eqs.~\eqref{current3} and \eqref{current4}) are almost zero. Whereas for the negative dc voltage the phase is nearly $\pi/2$ and the current increases approximately linearly with the amplitude, $I\backsim A_{3}\varkappa e\omega_{0}$, where $\varkappa$ is a numerical factor of order one.

\begin{figure}
\includegraphics[width=8.5cm]{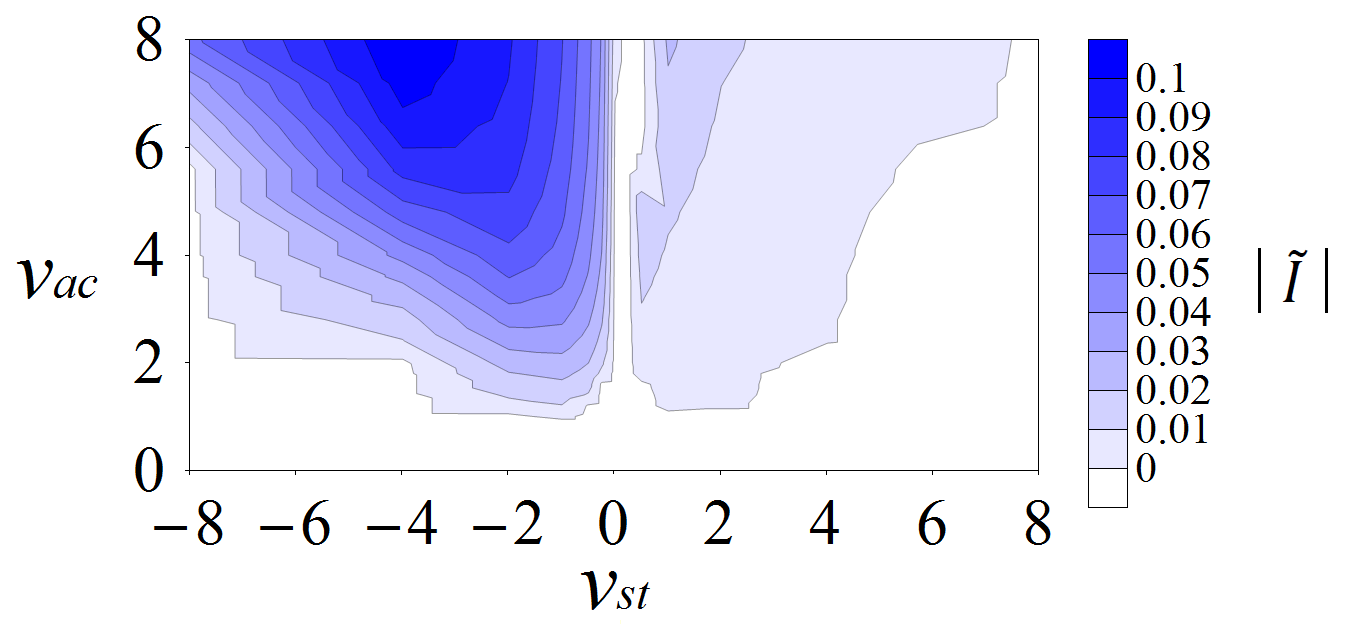}
\caption{(Color online) Dimensionless current $\vert \tilde{I} \vert$ obtained by numerical integration of Eqs.\eqref{eq1} as a function of the applied voltages. Current through the nanostructure is significal for negative ac voltage values. In the plot $\tilde{\epsilon}=0.1$, $\alpha= 837$, $\tilde{\omega} = 1$.}
\label{corren}
\end{figure}

Finally, we investigated the current behavior as a function of the applied voltages $\tilde{I}=\tilde{I}(\nu_{st},\nu_{ac})$. The result is displayed in Fig.~\ref{corren}. From the contour map, it is evident that the conductive behavior of the system with respect to the stationary dc voltage $\nu_{st}$ is asymmetric. Therefore, in light of the result previously discussed, the values of the applied voltages should be choosen in order to maximize the charge transport. In this case, from Fig.~\ref{corriente}, we look for a stationary phase $\chi_{3}= \pm \pi/2$ and, from Fig.~3(a), this condition is more likely achieved for negative values of $\nu_{st}$, as shown in Fig.~\ref{corren}.

To conclude, we have analized  parametric effects in a nanoelectromechanical single electron device in the form of a three-terminal shuttle system under the influence of a gate voltage. In particular, we have shown that in the parametric excitation regime, \textit{i.e.}, when the frequency of the gate voltage is approximately equal to the mechanical frequency of the nano-oscillator, electrons are transferred between the electrodes maintained at the same potential. In this case, the direction of the shuttle current is controlled by the phase shift between the mechanical vibrations and gate voltage oscillations.

All authors contributed equally to this work. Financial support form the Swedish Research Council (VR) is gratefully acknowledged.


\begin{thebibliography}{15}
\expandafter\ifx\csname natexlab\endcsname\relax\def\natexlab#1{#1}\fi
\expandafter\ifx\csname bibnamefont\endcsname\relax
  \def\bibnamefont#1{#1}\fi
\expandafter\ifx\csname bibfnamefont\endcsname\relax
  \def\bibfnamefont#1{#1}\fi
\expandafter\ifx\csname citenamefont\endcsname\relax
  \def\citenamefont#1{#1}\fi
\expandafter\ifx\csname url\endcsname\relax
  \def\url#1{\texttt{#1}}\fi
\expandafter\ifx\csname urlprefix\endcsname\relax\def\urlprefix{URL }\fi
\providecommand{\bibinfo}[2]{#2}
\providecommand{\eprint}[2][]{\url{#2}}

\bibitem[{\citenamefont{Gorelik et~al.}(1998)\citenamefont{Gorelik, Isacsson,
  Voinova, Kasemo, Shekhter, and Jonson}}]{Gorelik}
\bibinfo{author}{\bibfnamefont{L.}~\bibnamefont{Gorelik}},
  \bibinfo{author}{\bibfnamefont{A.}~\bibnamefont{Isacsson}},
  \bibinfo{author}{\bibfnamefont{M.}~\bibnamefont{Voinova}},
  \bibinfo{author}{\bibfnamefont{B.}~\bibnamefont{Kasemo}},
  \bibinfo{author}{\bibfnamefont{R.}~\bibnamefont{Shekhter}}, \bibnamefont{and}
  \bibinfo{author}{\bibfnamefont{M.}~\bibnamefont{Jonson}},
  \bibinfo{journal}{Phys. Rev. Lett.} \textbf{\bibinfo{volume}{80}},
  \bibinfo{pages}{4526} (\bibinfo{year}{1998}).

\bibitem[{\citenamefont{Gorelik et~al.}(2001)\citenamefont{Gorelik, Isacsson,
  Galperin, Shekhter, and Jonson}}]{gorelik2}
\bibinfo{author}{\bibfnamefont{L.~Y.} \bibnamefont{Gorelik}},
  \bibinfo{author}{\bibfnamefont{A.}~\bibnamefont{Isacsson}},
  \bibinfo{author}{\bibfnamefont{Y.~M.} \bibnamefont{Galperin}},
  \bibinfo{author}{\bibfnamefont{R.~I.} \bibnamefont{Shekhter}},
  \bibnamefont{and} \bibinfo{author}{\bibfnamefont{M.}~\bibnamefont{Jonson}},
  \bibinfo{journal}{Nature} \textbf{\bibinfo{volume}{411}},
  \bibinfo{pages}{454} (\bibinfo{year}{2001}).

\bibitem[{\citenamefont{Shekhter et~al.}(2003)\citenamefont{Shekhter, Galperin,
  Gorelik, Isacsson, and Jonson}}]{Shekhter}
\bibinfo{author}{\bibfnamefont{R.~I.} \bibnamefont{Shekhter}},
  \bibinfo{author}{\bibfnamefont{Y.~M.} \bibnamefont{Galperin}},
  \bibinfo{author}{\bibfnamefont{L.~Y.} \bibnamefont{Gorelik}},
  \bibinfo{author}{\bibfnamefont{A.}~\bibnamefont{Isacsson}}, \bibnamefont{and}
  \bibinfo{author}{\bibfnamefont{M.}~\bibnamefont{Jonson}},
  \bibinfo{journal}{J. Phys.: Condens. Matter} \textbf{\bibinfo{volume}{15}},
  \bibinfo{pages}{R441} (\bibinfo{year}{2003}).

\bibitem[{\citenamefont{Armour and MacKinnon}(2002)}]{armour}
\bibinfo{author}{\bibfnamefont{A.~D.} \bibnamefont{Armour}} \bibnamefont{and}
  \bibinfo{author}{\bibfnamefont{A.}~\bibnamefont{MacKinnon}},
  \bibinfo{journal}{Phys. Rev. B} \textbf{\bibinfo{volume}{66}},
  \bibinfo{pages}{035333} (\bibinfo{year}{2002}).

\bibitem[{\citenamefont{Azuma et~al.}(2007)\citenamefont{Azuma, Hatanaka,
  Kanehara, Teranishi, Chorley, Prance, Smith, and Majima}}]{azuma}
\bibinfo{author}{\bibfnamefont{Y.}~\bibnamefont{Azuma}},
  \bibinfo{author}{\bibfnamefont{T.}~\bibnamefont{Hatanaka}},
  \bibinfo{author}{\bibfnamefont{M.}~\bibnamefont{Kanehara}},
  \bibinfo{author}{\bibfnamefont{T.}~\bibnamefont{Teranishi}},
  \bibinfo{author}{\bibfnamefont{S.}~\bibnamefont{Chorley}},
  \bibinfo{author}{\bibfnamefont{J.}~\bibnamefont{Prance}},
  \bibinfo{author}{\bibfnamefont{C.~G.} \bibnamefont{Smith}}, \bibnamefont{and}
  \bibinfo{author}{\bibfnamefont{Y.}~\bibnamefont{Majima}},
  \bibinfo{journal}{Appl. Phys. Lett.} \textbf{\bibinfo{volume}{91}},
  \bibinfo{pages}{053120} (\bibinfo{year}{2007}).

\bibitem[{\citenamefont{Koenig et~al.}(2008)\citenamefont{Koenig, Weig, and
  Kotthaus}}]{koenig}
\bibinfo{author}{\bibfnamefont{D.~R.} \bibnamefont{Koenig}},
  \bibinfo{author}{\bibfnamefont{E.~M.} \bibnamefont{Weig}}, \bibnamefont{and}
  \bibinfo{author}{\bibfnamefont{J.~P.} \bibnamefont{Kotthaus}},
  \bibinfo{journal}{Nature Nanotech.} \textbf{\bibinfo{volume}{3}},
  \bibinfo{pages}{482} (\bibinfo{year}{2008}).

\bibitem[{\citenamefont{Moskalenko et~al.}(2009)\citenamefont{Moskalenko,
  Gordeev, Koentjoro, Raithby, French, Marken, and Savel'ev}}]{mosalenko}
\bibinfo{author}{\bibfnamefont{A.~V.} \bibnamefont{Moskalenko}},
  \bibinfo{author}{\bibfnamefont{S.~N.} \bibnamefont{Gordeev}},
  \bibinfo{author}{\bibfnamefont{O.~F.} \bibnamefont{Koentjoro}},
  \bibinfo{author}{\bibfnamefont{P.~R.} \bibnamefont{Raithby}},
  \bibinfo{author}{\bibfnamefont{R.~W.} \bibnamefont{French}},
  \bibinfo{author}{\bibfnamefont{F.}~\bibnamefont{Marken}}, \bibnamefont{and}
  \bibinfo{author}{\bibfnamefont{S.~E.} \bibnamefont{Savel'ev}},
  \bibinfo{journal}{Phys. Rev. B} \textbf{\bibinfo{volume}{79}},
  \bibinfo{pages}{241403} (\bibinfo{year}{2009}).

\bibitem[{\citenamefont{Kim et~al.}(2010)\citenamefont{Kim, Park, and
  Blick}}]{kim}
\bibinfo{author}{\bibfnamefont{C.}~\bibnamefont{Kim}},
  \bibinfo{author}{\bibfnamefont{J.}~\bibnamefont{Park}}, \bibnamefont{and}
  \bibinfo{author}{\bibfnamefont{R.~H.} \bibnamefont{Blick}},
  \bibinfo{journal}{Phys. Rev. Lett.} \textbf{\bibinfo{volume}{105}},
  \bibinfo{pages}{067204} (\bibinfo{year}{2010}).

\bibitem[{\citenamefont{Azuma et~al.}(2011)\citenamefont{Azuma, Kobayashi,
  Chorley, Prance, Smith, Tanaka, Kanehara, Teranishi, and Majima}}]{azuman}
\bibinfo{author}{\bibfnamefont{Y.}~\bibnamefont{Azuma}},
  \bibinfo{author}{\bibfnamefont{N.}~\bibnamefont{Kobayashi}},
  \bibinfo{author}{\bibfnamefont{S.}~\bibnamefont{Chorley}},
  \bibinfo{author}{\bibfnamefont{J.}~\bibnamefont{Prance}},
  \bibinfo{author}{\bibfnamefont{C.~G.} \bibnamefont{Smith}},
  \bibinfo{author}{\bibfnamefont{D.}~\bibnamefont{Tanaka}},
  \bibinfo{author}{\bibfnamefont{M.}~\bibnamefont{Kanehara}},
  \bibinfo{author}{\bibfnamefont{T.}~\bibnamefont{Teranishi}},
  \bibnamefont{and} \bibinfo{author}{\bibfnamefont{Y.}~\bibnamefont{Majima}},
  \bibinfo{journal}{Appl. Phys. Lett.} \textbf{\bibinfo{volume}{109}},
  \bibinfo{pages}{024303} (\bibinfo{year}{2011}).

\bibitem[{\citenamefont{Isacsson}(2001)}]{isacsson}
\bibinfo{author}{\bibfnamefont{A.}~\bibnamefont{Isacsson}},
  \bibinfo{journal}{Phys. Rev. B} \textbf{\bibinfo{volume}{64}},
  \bibinfo{pages}{035326} (\bibinfo{year}{2001}).

\bibitem[{\citenamefont{Karabalin et~al.}(2009)\citenamefont{Karabalin, Cross,
  and Roukes}}]{karabalin}
\bibinfo{author}{\bibfnamefont{R.~B.} \bibnamefont{Karabalin}},
  \bibinfo{author}{\bibfnamefont{M.~C.} \bibnamefont{Cross}}, \bibnamefont{and}
  \bibinfo{author}{\bibfnamefont{M.~L.} \bibnamefont{Roukes}},
  \bibinfo{journal}{Phys. Rev. B} \textbf{\bibinfo{volume}{79}},
  \bibinfo{pages}{165309} (\bibinfo{year}{2009}).

\bibitem[{\citenamefont{Kenig et~al.}(2009{\natexlab{a}})\citenamefont{Kenig,
  Lifshitz, and Cross}}]{kenig}
\bibinfo{author}{\bibfnamefont{E.}~\bibnamefont{Kenig}},
  \bibinfo{author}{\bibfnamefont{R.}~\bibnamefont{Lifshitz}}, \bibnamefont{and}
  \bibinfo{author}{\bibfnamefont{M.~C.} \bibnamefont{Cross}},
  \bibinfo{journal}{Phys. Rev. E} \textbf{\bibinfo{volume}{79}},
  \bibinfo{pages}{026203} (\bibinfo{year}{2009}{\natexlab{a}}).

\bibitem[{\citenamefont{Kenig et~al.}(2009{\natexlab{b}})\citenamefont{Kenig,
  Malomed, Cross, and Lifshitz}}]{kenig1}
\bibinfo{author}{\bibfnamefont{E.}~\bibnamefont{Kenig}},
  \bibinfo{author}{\bibfnamefont{B.~A.} \bibnamefont{Malomed}},
  \bibinfo{author}{\bibfnamefont{M.~C.} \bibnamefont{Cross}}, \bibnamefont{and}
  \bibinfo{author}{\bibfnamefont{R.}~\bibnamefont{Lifshitz}},
  \bibinfo{journal}{Phys. Rev. E} \textbf{\bibinfo{volume}{80}},
  \bibinfo{pages}{046202} (\bibinfo{year}{2009}{\natexlab{b}}).

\bibitem[{\citenamefont{Midtvedt et~al.}(2011)\citenamefont{Midtvedt,
  Tarakanov, and Kinaret}}]{midtvedt}
\bibinfo{author}{\bibfnamefont{D.}~\bibnamefont{Midtvedt}},
  \bibinfo{author}{\bibfnamefont{Y.}~\bibnamefont{Tarakanov}},
  \bibnamefont{and} \bibinfo{author}{\bibfnamefont{J.}~\bibnamefont{Kinaret}},
  \bibinfo{journal}{Nano Lett.} \textbf{\bibinfo{volume}{11}},
  \bibinfo{pages}{1439} (\bibinfo{year}{2011}).

\bibitem[{prl()}]{prl_comment}
\bibinfo{note}{Here, we have assumed that chemical potential of the dot is
  equal to the chemical potential of the leads.}

\end{thebibliography}

\end{document}